\documentclass[10pt, prl,aps,twocolumn]{revtex4-1}
\usepackage{amssymb,amsmath,amsfonts,amsbsy,epsfig,graphicx}
\usepackage[dvipsnames]{xcolor}
\usepackage{braket}
\definecolor{lgray}{gray}{0.35}
\usepackage[colorlinks=true,urlcolor=Gray,linkcolor=lgray,
citecolor=Gray,
             pdfpagelabels=true,hypertexnames=true,
            plainpages=false,naturalnames=false,
             ]{hyperref}

\newcommand{\be}{\begin{equation}}
\newcommand{\ee}{\end{equation}}
\newcommand{\bea}{\begin{eqnarray}}
\newcommand{\eea}{\end{eqnarray}}
\newcommand{\nn}{\nonumber}

\newcommand{\dph}{{\dot{\phi}}}
\def\k{{\vec{k}}}
\def\q{{\vec{q}}}
\begin{document}

\title{Constraints on holographic multi-field inflation and models based on the Hamilton-Jacobi formalism}


\author{
Ana Ach\'ucarro$^{a,b}$, Sebasti\'an C\'espedes$^{c}$, Anne-Christine Davis$^{c}$, and Gonzalo A. Palma$^{d}$
} 

\affiliation{
$^{a}$Instituut-Lorentz for Theoretical Physics, Universiteit Leiden, \mbox{2333 CA Leiden, The Netherlands.} \\
$^{b}$Department of Theoretical Physics, University of the Basque Country, \mbox{48080 Bilbao, Spain.}\\
$^{c}$DAMTP, University of Cambridge, Wilberforce Road, Cambridge, CB3 0WA, UK.\\
$^{d}$Grupo de Cosmolog\'ia y Astrof\'isica Te\'orica, Departamento de F\'{i}sica, FCFM, \mbox{Universidad de Chile}, Blanco Encalada 2008, Santiago, Chile. 
}

\begin{abstract}

In holographic inflation, the $4D$ cosmological dynamics is postulated to be dual to the renormalization group flow of a $3D$ Euclidean conformal field theory with marginally relevant operators. The scalar potential of the $4D$ theory ---in which inflation is realized--- is highly constrained, with use of the Hamilton--Jacobi equations. In multi-field holographic realizations of inflation, fields additional to the inflaton cannot display underdamped oscillations (that is, their wavefunctions contain no oscillatory phases independent of the momenta). We show that this result is exact, independent of the number of fields, the field space geometry and the shape of the inflationary trajectory followed in multi-field space. In the specific case where the multi-field trajectory is a straight line or confined to a plane, it can be understood as the existence of an upper bound on the dynamical masses $m$ of extra fields of the form $m \leq 3 H / 2$ up to slow roll corrections. This bound corresponds to the analytic continuation of the well known Breitenlohner--Freedman bound found in AdS spacetimes in the case when the masses are approximately constant. The absence of underdamped oscillations implies that a detection of ``cosmological collider" oscillatory patterns in the non-Gaussian bispectrum would not only rule out single field inflation, but also holographic inflation or any inflationary model based on the Hamilton--Jacobi equations.  Hence, future observations have the potential to exclude, at once, an entire class of inflationary theories, regardless of the details involved in their model building.

\end{abstract}

\maketitle

\emph{Introduction}: The observation of departures from a perfectly Gaussian distribution of primordial curvature perturbations would allow us to infer fundamental information about cosmic inflation~\cite{Guth:1980zm, Starobinsky:1980te, Linde:1981mu, Albrecht:1982wi, Mukhanov:1981xt}. It is by now well understood that single field models of inflation cannot account for primordial local non-Gaussianity unless a nontrivial self-interaction, together with a non-attractor background evolution, plays a role in inducing it~\cite{Maldacena:2002vr, Tanaka:2011aj, Pajer:2013ana,  Cabass:2016cgp, Bravo:2017gct}. This is mostly due to the fact that the dynamics of curvature perturbations is highly constrained by the diffeomorphism invariance of the gravitational theory within which inflation is realized. On the other hand, multi-field inflation can accommodate non-gravitational interactions affecting the dynamics of curvature perturbations: fields orthogonal to the inflationary trajectory can efficiently transfer their non-Gaussian statistics ---resulting from their own self-interactions--- to curvature perturbations~\cite{Lyth:2005fi, Seery:2005gb, Rigopoulos:2005ae, Byrnes:2008wi, Chen:2009we, Chen:2009zp, Baumann:2011nk, Achucarro:2016fby, Chen:2018uul, Chen:2018brw}. As such, the detection of non-Gaussianity could reveal signatures only attributable to additional degrees of freedom interacting with curvature perturbations~\cite{Arkani-Hamed:2015bza, Dimastrogiovanni:2015pla, Mirbabayi:2015hva, Chen:2015lza, Flauger:2016idt, Lee:2016vti}. 

Understanding the theoretical restrictions on the various classes of interactions coupling together fields in multi-field systems would allow us to interpret future observations related to non-Gaussianity. For example, multi-field systems derived from supergravity, characterized by non-flat K\"ahler geometries, are severely restricted due to the way in which the gravitational interaction couples chiral fields together. As a consequence, it is not easy to spontaneously break supersymmetry and keep every chiral field stabilized while sustaining inflation. A similar situation holds in string theory compactifications, where many fields have a geometrical origin restricting their couplings at energies below the compactification scale, making it hard to build a quasi-de Sitter stage where all moduli are stabilized (see Ref.~\cite{Danielsson:2018ztv} and references therein). These restrictions do not only impose a challenge to the construction of realistic models of inflation, but they also have consequences for the prediction of observable primordial spectra~\cite{Hetz:2016ics}.

Other classes of well motivated multi-field constructions, enjoying a constrained structure, have received less attention. In particular, holographic models of inflation have interactions constrained by certain requirements on the ``holographic" correspondence connecting the $4D$ inflationary bulk cosmology and the $3D$ Euclidean conformal field theory (CFT)~\cite{Larsen:2002et, vanderSchaar:2003sz, Schalm:2012pi, Bzowski:2012ih, Mata:2012bx,McFadden:2013ria, Garriga:2014fda}. In these theories, the $4D$ dynamics is dual to a renormalization group flow realized in a strongly coupled $3D$ Euclidean CFT with marginally relevant scalar operators deforming the conformal symmetry. The action determining the 4D dynamics is conjectured to be
\be
S = \!\! \int \! d^4 x \sqrt{-g} \left[ \frac{1}{2} R - \frac{1}{2} \gamma_{ab} (\phi)  \, \partial_{\mu} \phi^a \partial^{\mu} \phi^b - V (\phi) \right] ,
\ee
where $R$ is the Ricci scalar constructed from the spacetime metric $g_{\mu\nu}$ (in units where the reduced Planck mass is $1$), and $\gamma_{a b}$ is a sigma-model metric characterizing the geometry of the multi-scalar field space spanned by $\phi^a$ (with $a = 1, \cdots, 1+N$). The potential $V$ is determined by a ``fake" superpotential $W(\phi)$ as
\be
V = 3 W^2 - 2 \gamma^{a b} W_a W_b ,  \label{potential_V-W}
\ee
where $\gamma^{ab}$ is the inverse of $\gamma_{a b}$, and $W_a = \partial W / \partial \phi^a$.  The inflationary solutions admitted by (\ref{potential_V-W}) that are dual to the renormalization group flow are given by Hamilton--Jacobi equations, of the form 
\be
\dot \phi^a = - 2 \gamma^{a b} W_b ,  \qquad\qquad H = W , \label{inflationary_path} 
\ee
where $H = \dot a / a$ is the Hubble parameter. The trajectory described by this solution is dual to the renormalization group flow of the boundary operators with fixed points representing static de Sitter configurations of the cosmological bulk. Thus, the entire cosmological history, starting from a static de Sitter universe (inflation), and ending in another static de Sitter universe (our dark energy dominated universe) may be understood as the consequence of renormalization group flow from the UV-fixed point (late universe) to the IR-fixed point (early universe). Another  class of  holographic models where a non-geometric 4D holographic spacetime is  associated with a weakly-coupled 3D CFT was studied in \cite{McFadden:2009fg}, and its  observational consequences in~\cite{Afshordi:2016dvb}. 

The purpose of this letter is to study some of the consequences on the dynamics of multi-field fluctuations coming from the constrained structure of the potential in eq.~(\ref{potential_V-W}). Our goal is to understand how the structure of~(\ref{potential_V-W}), together with~(\ref{inflationary_path}), constrains the interactions between the primordial curvature perturbation and other (isocurvature) fields during inflation and, more importantly, how this affects their observation. Our results apply to any model described by the Hamilton--Jacobi equations (\ref{potential_V-W}) and (\ref{inflationary_path}), regardless of the holographic interpretation. Our analysis will revolve around a known upper mass bound on all fields additional to the inflaton (as well as on the inflaton) given by
\be
m \leq m_{\rm max } \equiv 3H/2 .  \label{bound-1}
\ee
This bound was derived in \cite{Skenderis:2006rr} in the single-field case, and argued to be valid in the multifield case in \cite{Skenderis:2006jq} \emph{under the implicit assumption that all masses are constant}. It coincides with the analytic continuation of the Breitenlohner--Freedman bound encountered in scalar field theories in AdS spacetimes~\cite{Breitenlohner:1982bm}. The main consequence emerging from (\ref{bound-1}) is that fluctuations are forbidden to display under-damped oscillations.

As we shall see, in general multi-field holographic inflation, both sides of (\ref{bound-1}) receive corrections. First of all, for fields orthogonal to the trajectory, the bound applies to the dynamical, ``entropy" mass matrix, which differs from that obtained from the Hessian of the potential. Secondly, the upper bound receives corrections if the masses evolve in time, which is the generic situation during inflation. Assuming slow roll, the bound (\ref{bound-1}) receives small deformations in the cases where the trajectory is a straight line, or if it is confined to a plane (even with strong bending rates). In more general situations (for example, a spiraling path), the structure of the entropy mass-matrix becomes highly nontrivial and the generalization of eq. (\ref{bound-2}) is not very illuminating. Nevertheless, one can focus on the  evolution of fluctuations to show that \emph{fields additional to the inflaton will not have underdamped oscillations, regardless of the number of fields, the field space geometry and/or the shape of the inflationary trajectory followed in multi-field space}. Because fields with underdamped oscillations lead to distinguishable non-Gaussian features that could be observed in future surveys, their observation would immediately rule out holographic versions of inflation or any other model based on the multi-field Hamilton-Jacobi equations (\ref{inflationary_path}).

\emph{Derivation of the bound}:  We can motivate the bound by considering the simplest case: a straight trajectory in a model with $1+N$ fields with canonical kinetic terms $\gamma_{ab} = \delta_{ab}$. In this case the dynamical ``entropy" mass coincides with the naive mass (Hessian of $V$). Without loss of generality, we can take the inflationary trajectory along the $\phi^{N+1} \equiv \phi$ direction, with all other fields stabilized: $\phi^i \equiv \sigma^i = \sigma^i_0$ for $i = 1 , \cdots , N$. Note that (\ref{inflationary_path}) implies $W_{\sigma^i} = 0$ on the inflationary trajectory, and we can expand the superpotential as $W= w (\phi) +  \frac{1}{2} \sum_{i=1}^N  h_i  (\phi) (\sigma^i - \sigma_0^i)^2 + \cdots$, where $w(\phi)$ and $h_i (\phi)$ are given functions of $\phi$. Inserting this expression back into~(\ref{potential_V-W}) gives $V = 3 w^2 -2 (w ')^2 + \frac{1}{2}\sum_i m_i^2 (\phi) (\sigma^i - \sigma_0^i)^2$, where the masses $m_i(\phi)$ of the fields $\sigma^i$ are found to be given by $m_i^2 (\phi) = 6 w h_i - 4 h_i^2 - 4  w'  h'_i$. We can rewrite this expression in a more useful way by noticing from Eqs.~(\ref{inflationary_path}) that $w = H$ and $w' = - \dot \phi / 2$:
\be
m_i^2 = 6 H h_i  - 4 h_i^2  + 2 \dot h_i . \label{m-h}
\ee
Notice that for a constant $h_i$, the field $\sigma_i$ is non-tachyonic ($m_i^2>0$) as long as $0 < h_i < 3H/2$. Because in eq.~(\ref{m-h}) $m_i^2$ is a quadratic function of $h_i$, with a negative quadratic term, one obtains the following bound
\be 
m_i <  m_{\rm max} (1 + \delta_i /3)  ,  \label{bound-2}
\ee
where we have defined $\delta_i =  \dot h_i / H h_i$. Notice that $\delta_i$ measures the running of $h_i$. If background quantities evolve slowly, then we expect $\delta \sim \mathcal O(\epsilon)$, implying that masses stay almost constant during slow roll, and that the bound cannot be violated. If $\delta_i$ is large (of order 1) the field $h_i$, which near the maximum satisfies $h_i\sim 3H/4$, will typically evolve outside the nontachyonic domain within a few $e$-folds (unless $h_i \ll H$, in which case the value of the mass is far from the bound). For instance, suppose that we wanted to fix $m_i$ to a constant value $m_0$ larger than $m_{\rm max} = 3H/2$. Then (\ref{bound-2}) may be read as a differential equation for $h_i$ with a solution of the form
\be
h_i (t) =\frac{m_{\rm max}}{2}  + \frac{\Delta m}{2} \tan \left[ \Delta  m (t-t_0) \right] ,
\ee
where $\Delta  m \equiv \sqrt{m_0^2 - m_{\rm max}^2}$. This shows that $m_0$ can be larger than $m_{\rm max}$ but only for a very limited amount of time, which in $e$-folds is given by $\Delta N \sim H / \Delta  m$, thus making it impossible to achieve stable configurations where the masses stay above the bound $m_{\rm max}$ by a significant margin.

One way of understanding the emergence of the bound is as follows: Because the inflationary trajectory is dual to the renormalization group flow in the CFT side of the duality, the potential driving inflation must always admit monotonic solutions of the form (\ref{inflationary_path}), regardless of the initial conditions. This is satisfied for flows that are solutions of the Hamilton--Jacobi equations, which are monotonic in the sense that a trajectory satisfying (\ref{inflationary_path}) can never go back to a point already traversed (as they are gradient flows of the superpotential $W$). This notion coincides with the standard definition of monotonicity in the case single field models. This restricts the value of the masses of the fields, simply because a field with mass larger than $3 H /2$ allows for non-monotonic trajectories. To appreciate this, let us disregard the motion of the inflaton $\phi$ and focus on the background evolution of one of the massive fields $\sigma$ with a mass $m$. Its background equation of motion is given by 
\be
\ddot \sigma + 3 H \dot \sigma + m^2 (\sigma- \sigma_0) = 0. \label{eom-chi}
\ee 
The general solution is of the form $\sigma(t) = \sigma_0 +  A_{+} e^{\omega_{+} t} + A_{-} e^{\omega_{-} t}$ with:
\be
\omega_{\pm} = - \frac{3}{2} H  \pm \frac{3}{2} H \sqrt{ 1 - \frac{4}{9} \frac{m^2}{H^2}  } . \label{freqs}
\ee
If $m < m_{\rm max}$, the solutions  are overdamped, and the field $\sigma$ reaches $\sigma = \sigma_0$ monotonically at a time $t \gg H^{-1}$. On the other hand, if $m > m_{\rm max}$ the underdamped solutions are oscillatory, and not of the desired form $\dot \sigma = f(\sigma)$. Moreover, notice that by inserting the expression (\ref{m-h}) for $m^2$, with $\dot h =0$ back in (\ref{freqs}) the frequencies become
\be
\omega_{-} = - 2 h , \qquad  \omega_{+} = + 2 h - 3 H, \label{freqs-h}
\ee 
from here we see again that $0 < h_i < 3H/2$ is required so that the trajectory remains stable. Independently of this, eq.~(\ref{freqs-h}) shows us that, regardless of the value of $h$, the solutions are monotonic, and so the field cannot oscillate about the equilibrium point $\sigma_0$. 

\emph{Long-wavelength behavior of fluctuations:} The previous explanation helps to understand the origin of the bound affecting a massive field in a de Sitter spacetime, and mild deformations of it, such as the case of a straight inflationary trajectory in multi-field space. But, as could be expected, in multi-field models with arbitrarily bending trajectories, the deformations to the bound can be substantial. In what follows we revisit the previous discussion in the most general case, where $\gamma_{a b}$ is non-canonical and the inflationary trajectory in multi-field space does not correspond to a straight line. To start with, it is convenient to anticipate a few results. First, from eq.~(\ref{inflationary_path}) we see that if $W$ is a differentiable function of the fields, then it necessarily gives us back a unique set of background solutions. That is, provided an initial condition $\phi^a (t_0)$, there is only one possible solution $\phi^a(t)$ for $t>t_0$. This implies that two paths respecting~(\ref{inflationary_path}) can never cross each other, simply because the crossing point would constitute an initial condition yielding two different solutions. Let us consider one of such background solutions, $\phi_0^a(t)$, and perturb it. In the long-wavelength limit, where we can neglect its spatial dependence, the perturbed solution can be written as $\phi^a(t) = \phi_0^a(t) + \delta \phi^a(t)$. Now, given that $\phi^a(t)$ is independent of the spatial coordinates, there must exist some set of initial conditions for $\delta \phi^a$ such that $\phi^a(t)$ satisfies (\ref{inflationary_path}). In that case, $\delta \phi^a(t)$ must necessarily respect a first order differential  equation restricting its time evolution. To derive it, it is enough to expand (\ref{inflationary_path}) around $\phi^a(t) =  \phi_0^a(t) + \delta \phi^a(t)$. One finds
\be
\left[ \gamma_{ab} D_t  + 2 \nabla_a \nabla_b W  \right]_0 \delta \phi^b = 0 , \label{first-order-pert-eq}
\ee
where $D_t$ is a covariant derivative defined to act on vectors as $D_t A^a = \dot A^a + \Gamma^{a}_{b c} \dot \phi^b A^c$, where $\Gamma^{a}_{bc}$ represents Christoffel symbols. Of course, even though the backgrounds under study satisfy first order differential equation, their perturbations must respect a second order differential equation of motion. However, the analysis leading to (\ref{first-order-pert-eq}) shows that on long-wavelengths there must exist at least one solution $\delta  \phi_0^a$ satisfying a first order homogeneous differential equation. We will come back to this result in brief.

\emph{Bending trajectories in arbitrary field geometries:} We now consider the dynamics of fluctuations in the most general situation possible. We introduce a local ``Frenet-Serret" frame of $1+N$ unit-vectors on the background inflationary path. The first vector $T^a$ is defined to be tangential to the trajectory $\phi^a = \phi^a (t)$
\be
T^a =  \dph_0^a / \dot \phi_0,
\ee
whereas the rest of the vectors are denoted as $U^a_I (t)$ with $I = 1 , \cdots, N$, and are defined to satisfy~\cite{GrootNibbelink:2001qt, Achucarro:2010da}:
\be
D_t U^a_I = \Omega_{I-1} U_{I-1}^a - \Omega_{I} U_{I+1}^a . \label{Dt-U}
\ee
(also valid for $T^a$ if one takes $U_0^a = T^a$, and $\Omega_{-1} = 0$).  $\Omega_{I-1}$ is the angular velocity describing the rate at which $U_{I}$ rotates into the direction $U_{I-1}$. For instance, $\Omega_0$ is the angular velocity with which $T^a$ rotates towards the normal direction $U_1^a$. It is useful to define the following antisymmetric matrix (only valid for $I,J \geq 1$)
\bea
 A_{IJ} &=&  \Omega_I \delta_{I (J-1)} - \Omega_J \delta_{(I-1)J} . \label{AIJ}
\eea
To study the dynamics of the inflationary fluctuations we derive the action of the fluctuations in co-moving gauge, with the perturbed metric given by $ds^2 = - dt^2 + a^2 e^{2 \zeta} d{\bf x}^2$, where $\zeta$ is the co-moving curvature perturbation~\cite{footnote2}.  On the other hand, the field-fluctuations $\delta \phi^a \equiv \phi^a - \phi_0^a$ can be parametrized in terms of  isocurvature $\psi_I$ fields as $\delta \phi^a = \sum_I U^a_I \psi_I$~\cite{Gordon:2000hv}. Here $\psi_I$ corresponds to a fluctuation along the direction $U^a_I$. Notice that in this gauge the fluctuation along $T^a$ is set to vanish. The quadratic action is found to be:
\bea
S &=& \frac{1}{2} \int d^4 x a^3 \bigg[ 2 \epsilon \left( \dot \zeta - \frac{2 \Omega_0}{\sqrt{2 \epsilon}} \psi_1 \right)^2 - \frac{2 \epsilon}{a^2} (\nabla \zeta)^2 \nn \\
&& +  \left( D_t \vec \psi \right)^2 + \frac{1}{a^2} (\nabla \vec \psi)^2 + \vec \psi\,^T \!\! \cdot \! M^2 \! \cdot \! \vec \psi   \bigg] , \label{action-psi}
\eea
where $D_t \vec \psi = d \vec \psi / dt +  A \vec \psi$, and $A$ is the matrix defined in (\ref{AIJ}). The entropy mass matrix $M^2$ is given by 
\be
M_{IJ}^2 = V_{IJ} + \dot \phi_0^2  \mathbb{R}_{I J} + 3 \Omega_0^2 \delta_{1I}\delta_{1J},
\ee
where $V_{IJ} \equiv U_I^a U_J^b \nabla_a V_b$, and $ \mathbb{R}_{I J}  \equiv T^a U^b_I T^c U^d_I \mathbb{R}_{abcd}$ (with $\mathbb{R}_{abcd}$ the Riemann tensor associated to  $\gamma_{ab}$). Notice that the entropy mass matrix differs from the Hessian of the potential. In particular, it receives a contribution from the curvature tensor $ \mathbb{R}_{I J}$ (whose effect has been studied in~\cite{Renaux-Petel:2015mga, Cicoli:2018ccr}), and the angular velocity $\Omega_0$. One can now perform a field redefinition to a new frame where the isocurvature fields are canonical~\cite{Achucarro:2010da}. This is achieved by the following rotation $\vec \sigma = R(t) \vec \psi$, where $R(t) = \mathcal T e^{\int^t \!\! A(t)}$, with $\mathcal T$ the usual time ordering symbol. This rotation matrix keeps track of the bending of the trajectory [recall the meaning of $\Omega_I$ in eq.~(\ref{AIJ})], and implies $d \vec \sigma / dt = R(t) D_t \vec \psi$. In the (canonical) $\sigma$-frame the mass matrix is  $\bar M^2 = R(t) M^2 R^T(t)$. In the long-wavelength limit  $\zeta$ can be solved in terms of $\vec \sigma$, giving
\be
\ddot {\vec \sigma} + 3 H  \dot {\vec \sigma} + \bar M^2 \vec \sigma = 0 . \label{eq-sigma-I}
\ee
This is the multi-field analogue of eq.~(\ref{eom-chi}). The advantage of working with $\vec \sigma$ (instead of $\vec \psi$) is that the kinetic terms of different components remain decoupled. However, for $\Omega_I \neq 0$ the mass matrix $\bar M^2$ can have a strong time dependence, as opposed to $M^2$, which evolves slowly.

Up until now eq.~(\ref{action-psi}) is completely general, and it assumes nothing about $V$. To study the long-wavelength behavior of holographic systems it is useful to define the Hessian of $W$ along the $U$-basis as $W_{IJ}  \equiv U_I^aU_J^b\nabla_aW_b$. Then, using (\ref{inflationary_path}) it is straightforward to find the following results for the projection along $T^a$: $W_{00} = \frac{1}{4}H(2\epsilon-\eta)$, and $W_{0I} =\frac{1}{2}\Omega_0 \delta_{1 I}$, where $\epsilon \equiv - \dot H / H^2$ and $\eta \equiv \dot \epsilon / H \epsilon$ are the usual slow-roll parameters, assumed to be small.Then, a tedious but straightforward computation leads to the following expression for the mass matrix $\bar M^2$:
\be
\bar M_{IJ}^2 = 6 H \bar W_{IJ} - 4 \bar W_{IK} \bar W_{KJ} + 2 \dot {\bar W}_{IJ} , \label{barMIJ} 
\ee
where $\bar W_{IJ} = R_{IK}(t) W_{KL} R_{LJ}^T(t)$. This is one of our main results. Notice that $\bar M_{IJ}^2$ has precisely the same structure as Eq. (\ref{m-h})  for the masses of fields along straight trajectories, where $h_i$ played the role of the Hessian $\bar W_{IJ}$. 

Given that $\Omega_0$ does not enter the definition of $A_{IJ}$ or $R(t)$, one immediately sees that if $N=1$ (two field models) or only $\Omega_0 \neq 0$ (planar trajectories), ${\bar W}_{IJ}$ evolves slowly and $\dot {\bar W}_{IJ}$ is slow-roll suppressed. Then, by diagonalizing $W_{IJ}$, one recovers the universal bound (\ref{bound-2}) on the eigenvalues of $\bar M_{IJ}^2$. It would be tempting to conclude that this is true in more general situations, where all $\Omega_I$'s are nonvanishing, but this is not possible. The structure displayed by (\ref{barMIJ}) is very constrained, but it doesn't lead to a simple universal bound on its eigenvalues (because ${\bar W}_{IJ}$ and $\dot {\bar W}_{IJ}$ cannot be diagonalized simultaneously). However, given that $W_{IJ}$ and $\bar W_{IJ}$ share the same eigenvalues, by taking the trace of eq.~(\ref{barMIJ}) it is direct to show that each eigenvalue of $\bar M_{IJ}^2$ is bounded above --up to slow roll corrections analogous to the $\delta_i$ terms in eq. (\ref{bound-2})-- by threshold values $m_{{\rm max} \, I}^2$ satisfying 
\be\frac{1}{N}\sum_{I=1}^{N} m_{{\rm max} \, I}^2 = m_{\rm max}^2 \ .
\ee
This means that in the general case of non-planar, strongly turning (slow roll) trajectories the threshold values $m_{{\rm max} \, I}^2$ split around $m_{\rm max}^2$ and this bound is not very useful in practice.

On the other hand, instead of pursuing explicit expressions for $m_{{\rm max} \, I}^2$, we can directly compute the wavefunctions $\sigma_I$ in the longwavelength limit. Indeed, the very particular form of the mass matrix (\ref{barMIJ}) allows for a ``BPS" factorization of eq.~(\ref{eq-sigma-I}): 
\be
\left[ \frac{d}{dt} - 2 \bar W + 3 H  \right] \left[  \frac{d}{dt} + 2 \bar W\right]  \vec \sigma = 0, \label{factor-eq}
\ee
where $\bar W$ stands for the Hessian $\bar W_{IJ} (t)$. The rightmost parenthesis in~(\ref{factor-eq}) is the same differential operator as in eq.~(\ref{first-order-pert-eq}), but written in the $\sigma$-frame. Eq.~(\ref{factor-eq}) confirms our expectation that one of the long-wavelength modes must respect~(\ref{first-order-pert-eq}). The general solution of eq.~(\ref{factor-eq}) is
\be
\vec \sigma (t) = \mathcal T e^{-2\int^t \! \bar W} \left[ \vec C_1 +  \int^t \!\!\! dt' \,  \mathcal T e^{\int^{t'} \! ( 4 \bar W - 3 H)} \vec C_2 \right] , \label{solution}
\ee
where $\vec C_1$ and $\vec C_2$ are integration constants set by initial conditions. The term proportional to $\vec C_1$ corresponds to the long-wavelength solution that solves~(\ref{first-order-pert-eq}) in the $\sigma$-frame, whereas the term proportional to $\vec C_2$ corresponds to the second solution. In fact these two modes are the multi-field generalizations of the two modes in eq.~(\ref{freqs-h}). Similarly, we see that the trajectory is stable as long as the eigenvalues of $W_{IJ}$ are positive and smaller than $3H/2$ [analogous to the condition $h_i < 3H/2$ found after eq.~(\ref{m-h}) to avoid tachyons]. The salient point of this result is that the long-wavelength evolution of the perturbations is overdamped: given that the eigenvalues of $\bar W_{IJ}$ are real, there are no oscillatory phases present in $\vec \sigma (t)$.

\emph{Non-Gaussianity:} Let us now address the observational consequences of our previous result. Correlation functions for single field inflation are highly constrained by dilations and special conformal transformations, which are non linearly realized by $\zeta$ at horizon crossing~\cite{Creminelli:2004yq, Cheung:2007sv, Creminelli:2012ed, Hinterbichler:2012nm, Bravo:2017wyw}, particularly in the squeezed limit of the $3$-point function $\langle\zeta^3\rangle$, which is when one of the momenta is taken to be soft (much smaller in magnitude than the other two). However, isocurvature fields interacting with the curvature perturbation $\zeta$ during inflation can leave traces of their existence by enhancing the amplitude of non-Gaussianity up to levels that can be distinguished from single field models~\cite{Chen:2010xka}. For instance, it has been shown that if the masses of isocurvature fields are large enough, these will lead to oscillatory footprints in the shape of the bispectrum in momentum space~\cite{Arkani-Hamed:2015bza, Mirbabayi:2015hva, Dimastrogiovanni:2015pla, Chen:2015lza}. This prediction has been worked out for the particular case where the massive fields are weakly coupled to $\zeta$. In the language of the present letter, this corresponds to the case where the $\Omega_I/H$ are small (and  $\dot {\bar W}$ can be neglected in eq.~(\ref{barMIJ})). To be concrete, consider a single isocurvature field ($N=1$) with $\Omega_0/H \ll 1$. The 3-point function can be easily computed using the in-in formalism, in which case the interaction picture Hamiltonian induced by a non-vanishing $\Omega_0$ is given by $H_I(t) = - \int d^3x \left[  \mathcal L_{(2)}^{\rm int} + \mathcal L_{(3)}^{\rm int} \right]$ where:
\be
\mathcal L_{(2)}^{\rm int} \propto \Omega_0 \times  \dot\zeta\sigma   ,  \qquad  \mathcal L_{(3)}^{\rm int} \propto  \Omega_0 \times  \dot\zeta^2\sigma .
\label{InteractionHamiltonians}
\ee
The vertex $\mathcal L_{(3)}^{\rm int}$ induces an interaction between the curvature mode $\zeta$ and the massive field $\sigma$ leading to corrections to the zeroth order prediction for $\langle \zeta^3 \rangle$. In particular, the squeezed limit acquires a dependence on the mass of $\sigma$~\cite{Chen:2009zp, Arkani-Hamed:2015bza, Dimastrogiovanni:2015pla, Chen:2015lza, Mirbabayi:2015hva, Flauger:2016idt, Lee:2016vti} that can be summarized as follows: When $m <3H/2$, the fluctuation $\sigma$ experiences overdamped oscillations at horizon crossing, and one finds 
 \be
\langle\zeta_{\q}\zeta_{\k_1}\zeta_{\k_2}\rangle_{\sigma} \sim P_\zeta(q)P_\zeta(k)\left(\frac{q}{k}\right)^{\frac{3}{2}-\nu}  ,
\ee
where $\nu\equiv\sqrt{9/4- m^2/H^2}>0$, and  $\q$ is the soft momentum, such that $k_1\sim k_2\gg q$. On the other hand, for masses $m >3H/2$, oscillations are underdamped, and the bispectrum becomes
\be
\langle\zeta_{\q}\zeta_{\k_1}\zeta_{\k_2}\rangle_{\sigma} \sim P_\zeta(q)P_\zeta(k)\left(\frac{q}{k}\right)^{3/2} \!\!\! \cos(-i\nu\log\frac{q}{k}-\phi_0),
\label{cosmocollidersqueezedlimit}
\ee
where this time $\nu\equiv\sqrt{m^2/H^2-9/4}$, and the phase $\phi_0$ is fixed in terms of $\nu$. Now, the gist of the previous prediction is that the isocurvature fields can only create an interference pattern on the non-Gaussian statistics of $\zeta$~\cite{Arkani-Hamed:2015bza, Mirbabayi:2015hva} if they experience underdamped oscillations at horizon crossing. In the more general case, regardless of how complicated the couplings between $\zeta$ and $\sigma_I$ may be, the mere fact that the $\sigma$-fields do not show underdamped oscillations precludes them from leaving oscillatory footprints in the spectra. As a result, detecting signals such as that of (\ref{cosmocollidersqueezedlimit}) would rule out holographic models and any model based on the equations (\ref{potential_V-W}) and (\ref{inflationary_path}). This pattern is part of what is known as cosmological collider \cite{Arkani-Hamed:2015bza} signatures, and they could be observed with future surveys  by looking, for example,  at the dark matter distribution or the $21$~cm line~\cite{Chen:2016vvw,Abazajian:2016yjj,Meerburg:2016zdz}.

\emph{Concluding remarks:} Future cosmological surveys, aimed at characterizing the distribution of primordial curvature perturbations, will be able to constrain holographic realizations of inflation. As discussed elsewhere, the presence or absence of underdamped oscillations in the spectra of the theory crucially determines the shape of non-Gaussian imprints in the primordial distribution of curvature perturbations. If future observations reveal the existence of oscillatory features in the spectra (usually interpreted as the presence of massive fields with masses $m> 3H/2$), any multi-field model which uses the Hamilton--Jacobi equations, including holographic ones, would be ruled out.

\begin{acknowledgments}

\emph{Acknowledgments}: We wish to thank Jan Pieter van der Schaar, Kostas Skenderis and Yvette Welling for useful discussions and comments. AA acknowledges support by the Netherlands Organization for Scientific Research (NWO), the Netherlands  Organization for Fundamental Research in Matter (FOM), the Basque Government (IT-979-16) and the Spanish Ministry MINECO (FPA2015-64041-C2-1P).  SC is supported by CONICYT through its program Becas Chile and the Cambridge Overseas Trust. ACD is partially supported by STFC under grants No. ST/L000385/1 and ST/L000636/1. GAP acknowledges support from the Fondecyt Regular project number 1171811 (CONICYT).  ACD  and SC thank the University of Chile at Santiago for hospitality. GAP ans SC wish to thank the Instituut-Lorentz for Theoretical Physics, Universiteit Leiden for hospitality. AA and ACD thank the George and Cynthia Mitchell Foundation for hospitality at the Brinsop Court workshop. 

\end{acknowledgments}


\begin{thebibliography}{99}

\bibitem{Guth:1980zm} 
  A.~H.~Guth,
  ``The Inflationary Universe: A Possible Solution to the Horizon and Flatness Problems,''
  Phys.\ Rev.\ D {\bf 23}, 347 (1981).
  
\bibitem{Starobinsky:1980te} 
  A.~A.~Starobinsky,
  ``A New Type of Isotropic Cosmological Models Without Singularity,''
  Phys.\ Lett.\ B {\bf 91}, 99 (1980).
  
\bibitem{Linde:1981mu} 
  A.~D.~Linde,
  ``A New Inflationary Universe Scenario: A Possible Solution of the Horizon, Flatness, Homogeneity, Isotropy and Primordial Monopole Problems,''
  Phys.\ Lett.\ B {\bf 108}, 389 (1982).
  
\bibitem{Albrecht:1982wi} 
  A.~Albrecht and P.~J.~Steinhardt,
  ``Cosmology for Grand Unified Theories with Radiatively Induced Symmetry Breaking,''
  Phys.\ Rev.\ Lett.\  {\bf 48}, 1220 (1982).
  
\bibitem{Mukhanov:1981xt} 
  V.~F.~Mukhanov and G.~V.~Chibisov,
  ``Quantum Fluctuation and Nonsingular Universe. (In Russian),''
  JETP Lett.\  {\bf 33}, 532 (1981)
  [Pisma Zh.\ Eksp.\ Teor.\ Fiz.\  {\bf 33}, 549 (1981)].

\bibitem{Maldacena:2002vr} 
  J.~M.~Maldacena,
  ``Non-Gaussian features of primordial fluctuations in single field inflationary models,''
  JHEP {\bf 0305}, 013 (2003)
  [astro-ph/0210603].
  
\bibitem{Tanaka:2011aj} 
  T.~Tanaka and Y.~Urakawa,
  ``Dominance of gauge artifact in the consistency relation for the primordial bispectrum,''
  JCAP {\bf 1105}, 014 (2011)
  [arXiv:1103.1251 [astro-ph.CO]].

\bibitem{Pajer:2013ana} 
  E.~Pajer, F.~Schmidt and M.~Zaldarriaga,
  ``The Observed Squeezed Limit of Cosmological Three-Point Functions,''
  Phys.\ Rev.\ D {\bf 88}, no. 8, 083502 (2013)
  [arXiv:1305.0824 [astro-ph.CO]].
  
\bibitem{Cabass:2016cgp} 
  G.~Cabass, E.~Pajer and F.~Schmidt,
  ``How Gaussian can our Universe be?,''
  JCAP {\bf 1701}, no. 01, 003 (2017)
  [arXiv:1612.00033 [hep-th]].
  
\bibitem{Bravo:2017gct} 
  R.~Bravo, S.~Mooij, G.~A.~Palma and B.~Pradenas,
  ``Vanishing of local non-Gaussianity in canonical single field inflation,''
  JCAP {\bf 1805}, no. 05, 025 (2018)
  [arXiv:1711.05290 [astro-ph.CO]].

\bibitem{Lyth:2005fi} 
  D.~H.~Lyth and Y.~Rodriguez,
  ``The Inflationary prediction for primordial non-Gaussianity,''
  Phys.\ Rev.\ Lett.\  {\bf 95}, 121302 (2005)
  [astro-ph/0504045].

\bibitem{Seery:2005gb} 
  D.~Seery and J.~E.~Lidsey,
  ``Primordial non-Gaussianities from multiple-field inflation,''
  JCAP {\bf 0509}, 011 (2005)
  [astro-ph/0506056].

\bibitem{Rigopoulos:2005ae} 
  G.~I.~Rigopoulos, E.~P.~S.~Shellard and B.~J.~W.~van Tent,
  ``Large non-Gaussianity in multiple-field inflation,''
  Phys.\ Rev.\ D {\bf 73}, 083522 (2006)
  [astro-ph/0506704].

\bibitem{Byrnes:2008wi} 
  C.~T.~Byrnes, K.~Y.~Choi and L.~M.~H.~Hall,
  ``Conditions for large non-Gaussianity in two-field slow-roll inflation,''
  JCAP {\bf 0810}, 008 (2008)
  [arXiv:0807.1101 [astro-ph]].

\bibitem{Chen:2009we} 
  X.~Chen and Y.~Wang,
  ``Large non-Gaussianities with Intermediate Shapes from Quasi-Single Field Inflation,''
  Phys.\ Rev.\ D {\bf 81}, 063511 (2010)
  [arXiv:0909.0496 [astro-ph.CO]].

\bibitem{Chen:2009zp}
  X.~Chen and Y.~Wang,
  ``Quasi-Single Field Inflation and Non-Gaussianities,''
  JCAP {\bf 1004} (2010) 027
  [arXiv:0911.3380 [hep-th]].
  
\bibitem{Baumann:2011nk} 
  D.~Baumann and D.~Green,
  ``Signatures of Supersymmetry from the Early Universe,''
  Phys.\ Rev.\ D {\bf 85}, 103520 (2012)
  [arXiv:1109.0292 [hep-th]].
  
\bibitem{Achucarro:2016fby}
  A.~Ach\'ucarro, V.~Atal, C.~Germani and G.~A.~Palma,
  ``Cumulative effects in inflation with ultra-light entropy modes,''
  JCAP {\bf 1702} (2017) no.02,  013
  [arXiv:1607.08609 [astro-ph.CO]].
  
\bibitem{Chen:2018uul} 
  X.~Chen, G.~A.~Palma, W.~Riquelme, B.~Scheihing Hitschfeld and S.~Sypsas,
  ``Landscape tomography through primordial non-Gaussianity,''
  arXiv:1804.07315 [hep-th].
  
\bibitem{Chen:2018brw} 
  X.~Chen, G.~A.~Palma, B.~Scheihing Hitschfeld and S.~Sypsas,
  ``Reconstructing the inflationary landscape with cosmological data,''
  arXiv:1806.05202 [astro-ph.CO].

\bibitem{Arkani-Hamed:2015bza} 
  N.~Arkani-Hamed and J.~Maldacena,
  ``Cosmological Collider Physics,''
  arXiv:1503.08043 [hep-th].
  
\bibitem{Dimastrogiovanni:2015pla} 
  E.~Dimastrogiovanni, M.~Fasiello and M.~Kamionkowski,
  ``Imprints of Massive Primordial Fields on Large-Scale Structure,''
  JCAP {\bf 1602}, 017 (2016)
  [arXiv:1504.05993 [astro-ph.CO]].

\bibitem{Mirbabayi:2015hva} 
  M.~Mirbabayi and M.~Simonovi\'c‡,
  ``Effective Theory of Squeezed Correlation Functions,''
  JCAP {\bf 1603}, no. 03, 056 (2016)
  [arXiv:1507.04755 [hep-th]].
  
\bibitem{Chen:2015lza} 
  X.~Chen, M.~H.~Namjoo and Y.~Wang,
  ``Quantum Primordial Standard Clocks,''
  JCAP {\bf 1602}, no. 02, 013 (2016)
  [arXiv:1509.03930 [astro-ph.CO]].

\bibitem{Flauger:2016idt} 
  R.~Flauger, M.~Mirbabayi, L.~Senatore and E.~Silverstein,
  ``Productive Interactions: heavy particles and non-Gaussianity,''
  JCAP {\bf 1710}, no. 10, 058 (2017)
  [arXiv:1606.00513 [hep-th]].
  
\bibitem{Lee:2016vti} 
  H.~Lee, D.~Baumann and G.~L.~Pimentel,
  ``Non-Gaussianity as a Particle Detector,''
  JHEP {\bf 1612}, 040 (2016)
  [arXiv:1607.03735 [hep-th]].

\bibitem{Danielsson:2018ztv} 
  U.~H.~Danielsson and T.~Van Riet,
  ``What if string theory has no de Sitter vacua?,''
  arXiv:1804.01120 [hep-th].

\bibitem{Hetz:2016ics} 
  A.~Hetz and G.~A.~Palma,
  ``Sound Speed of Primordial Fluctuations in Supergravity Inflation,''
  Phys.\ Rev.\ Lett.\  {\bf 117}, no. 10, 101301 (2016)
  [arXiv:1601.05457 [hep-th]].

\bibitem{Larsen:2002et}
  F.~Larsen, J.~P.~van der Schaar and R.~G.~Leigh,
  ``De Sitter holography and the cosmic microwave background,''
  JHEP {\bf 0204} (2002) 047
  [hep-th/0202127].
  
\bibitem{vanderSchaar:2003sz} 
  J.~P.~van der Schaar,
  ``Inflationary perturbations from deformed CFT,''
  JHEP {\bf 0401}, 070 (2004)
  [hep-th/0307271].

\bibitem{Mata:2012bx}
  I.~Mata, S.~Raju and S.~Trivedi,
  ``CMB from CFT,''
  JHEP {\bf 1307} (2013) 015
  [arXiv:1211.5482 [hep-th]].
  
\bibitem{Schalm:2012pi}
  K.~Schalm, G.~Shiu and T.~van der Aalst,
  ``Consistency condition for inflation from (broken) conformal symmetry,''
  JCAP {\bf 1303} (2013) 005
  [arXiv:1211.2157 [hep-th]].

\bibitem{Bzowski:2012ih}
  A.~Bzowski, P.~McFadden and K.~Skenderis,
  ``Holography for inflation using conformal perturbation theory,''
  JHEP {\bf 1304} (2013) 047
  [arXiv:1211.4550 [hep-th]].

\bibitem{McFadden:2013ria} 
  P.~McFadden,
  ``On the power spectrum of inflationary cosmologies dual to a deformed CFT,''
  JHEP {\bf 1310}, 071 (2013)
  [arXiv:1308.0331 [hep-th]].

\bibitem{Garriga:2014fda} 
  J.~Garriga, K.~Skenderis and Y.~Urakawa,
  ``Multi-field inflation from holography,''
  JCAP {\bf 1501}, no. 01, 028 (2015)
  [arXiv:1410.3290 [hep-th]].

\bibitem{McFadden:2009fg}
  P.~McFadden and K.~Skenderis,
  ``Holography for Cosmology,''
  Phys.\ Rev.\ D {\bf 81} (2010) 021301
  [arXiv:0907.5542 [hep-th]].
  
\bibitem{Afshordi:2016dvb}
  N.~Afshordi, C.~Coriano, L.~Delle Rose, E.~Gould and K.~Skenderis,
  ``From Planck data to Planck era: Observational tests of Holographic Cosmology,''
  Phys.\ Rev.\ Lett.\  {\bf 118} (2017) no.4,  041301
  [arXiv:1607.04878 [astro-ph.CO]].

\bibitem{Skenderis:2006rr}
  K.~Skenderis and P.~K.~Townsend,
  ``Hamilton-Jacobi method for curved domain walls and cosmologies,''
  Phys.\ Rev.\ D {\bf 74} (2006) 125008
  [hep-th/0609056].

\bibitem{Skenderis:2006jq}
  K.~Skenderis and P.~K.~Townsend,
  ``Hidden supersymmetry of domain walls and cosmologies,''
  Phys.\ Rev.\ Lett.\  {\bf 96} (2006) 191301
  [hep-th/0602260].

\bibitem{Breitenlohner:1982bm}
  P.~Breitenlohner and D.~Z.~Freedman,
  ``Positive Energy in anti-De Sitter Backgrounds and Gauged Extended Supergravity,''
  Phys.\ Lett.\  {\bf 115B} (1982) 197.

\bibitem{GrootNibbelink:2001qt} 
  S.~Groot Nibbelink and B.~J.~W.~van Tent,
  ``Scalar perturbations during multiple field slow-roll inflation,''
  Class.\ Quant.\ Grav.\  {\bf 19}, 613 (2002)
  [hep-ph/0107272].

\bibitem{Achucarro:2010da} 
  A.~Achucarro, J.~O.~Gong, S.~Hardeman, G.~A.~Palma and S.~P.~Patil,
  ``Features of heavy physics in the CMB power spectrum,''
  JCAP {\bf 1101}, 030 (2011)
  [arXiv:1010.3693 [hep-ph]].
  
\bibitem{footnote2} This variable is often called $\cal R$ but here we follow the notation of \cite{Maldacena:2002vr}.

\bibitem{Gordon:2000hv}
  C.~Gordon, D.~Wands, B.~A.~Bassett and R.~Maartens,
  ``Adiabatic and entropy perturbations from inflation,''
  Phys.\ Rev.\ D {\bf 63}, 023506 (2001)
  [astro-ph/0009131].
  
\bibitem{Renaux-Petel:2015mga} 
  S.~Renaux-Petel and K.~Turzy?ski,
  ``Geometrical Destabilization of Inflation,''
  Phys.\ Rev.\ Lett.\  {\bf 117}, no. 14, 141301 (2016)
  [arXiv:1510.01281 [astro-ph.CO]].

\bibitem{Cicoli:2018ccr}
  M.~Cicoli, V.~Guidetti, F.~G.~Pedro and G.~P.~Vacca,
  ``A geometrical instability for ultra-light fields during inflation?,''
  arXiv:1807.03818 [hep-th].

\bibitem{Creminelli:2004yq} 
  P.~Creminelli and M.~Zaldarriaga,
  ``Single field consistency relation for the 3-point function,''
  JCAP {\bf 0410}, 006 (2004)
  [astro-ph/0407059].
  
\bibitem{Cheung:2007sv}
  C.~Cheung, A.~L.~Fitzpatrick, J.~Kaplan and L.~Senatore,
  ``On the consistency relation of the 3-point function in single field inflation,''
  JCAP {\bf 0802}, 021 (2008)
  [arXiv:0709.0295 [hep-th]].

\bibitem{Creminelli:2012ed}
  P.~Creminelli, J.~Nore\~na and M.~Simonovi\'c‡,
  ``Conformal consistency relations for single-field inflation,''
  JCAP {\bf 1207} (2012) 052
  [arXiv:1203.4595 [hep-th]].

\bibitem{Hinterbichler:2012nm}
  K.~Hinterbichler, L.~Hui and J.~Khoury,
  ``Conformal Symmetries of Adiabatic Modes in Cosmology,''
  JCAP {\bf 1208} (2012) 017
  [arXiv:1203.6351 [hep-th]].

\bibitem{Bravo:2017wyw} 
  R.~Bravo, S.~Mooij, G.~A.~Palma and B.~Pradenas,
  ``A generalized non-Gaussian consistency relation for single field inflation,''
  JCAP {\bf 1805}, no. 05, 024 (2018)
  [arXiv:1711.02680 [astro-ph.CO]].

\bibitem{Chen:2010xka} 
  X.~Chen,
  ``Primordial Non-Gaussianities from Inflation Models,''
  Adv.\ Astron.\  {\bf 2010}, 638979 (2010)
  [arXiv:1002.1416 [astro-ph.CO]].

\bibitem{Meerburg:2016zdz}
  P.~D.~Meerburg, M. M\"unchmeyer, J.~B.~Mu\~noz and X.~Chen,
  ``Prospects for Cosmological Collider Physics,''
  JCAP {\bf 1703} (2017) no.03,  050
  [arXiv:1610.06559 [astro-ph.CO]].

\bibitem{Chen:2016vvw}
  X.~Chen, C.~Dvorkin, Z.~Huang, M.~H.~Namjoo and L.~Verde,
  ``The Future of Primordial Features with Large-Scale Structure Surveys,''
  JCAP {\bf 1611} (2016) no.11,  014
  [arXiv:1605.09365 [astro-ph.CO]].
  
\bibitem{Abazajian:2016yjj}
  K.~N.~Abazajian {\it et al.} [CMB-S4 Collaboration],
  ``CMB-S4 Science Book, First Edition,''
  arXiv:1610.02743 [astro-ph.CO].
  





\end{thebibliography}
\end{document}